\documentclass[10pt, conference, a4paper]{IEEEtranV1_8b}

\usepackage[utf8x]{inputenc}
\usepackage[english]{babel}
\usepackage[T1]{fontenc}
\usepackage{subfiles}
\usepackage{textcomp}
\usepackage[sort]{cite}
\usepackage{courier}
\usepackage{booktabs}
\usepackage{graphicx}
\usepackage{epstopdf}
\usepackage{float}
\usepackage[caption = false, font=footnotesize]{subfig}
\usepackage{moreverb}
\usepackage{color}
\usepackage{amsmath,amssymb,dsfont}
\usepackage{balance}
\usepackage{ifpdf}
\usepackage{multirow}
\usepackage{tabularx}
\usepackage{textcomp}
\usepackage{hhline}
\usepackage{algorithm,algorithmic}
\usepackage{enumerate}
\usepackage{amssymb}
\usepackage{capt-of}
\usepackage[%
	binary-units=true,
	per-mode=symbol,
	exponent-product=\cdot,
	range-units=single
	]{siunitx}
\usepackage{url}
\usepackage{listings}

\usepackage[absolute,showboxes]{textpos}

\usepackage[capitalise]{cleveref}

\newcolumntype{Z}{>{\centering\arraybackslash}X}

\makeatletter
\def\lst@makecaption{%
  \def\@captype{table}%
  \@makecaption
}
\makeatother

\begin{document}

\title{On Root Detection Strategies for Android Devices}

\author{\IEEEauthorblockN{Raphael Bialon}
\IEEEauthorblockA{Department of Computer Science, Heinrich-Heine-University Düsseldorf, Germany\\raphael.bialon@hhu.de}
}


\maketitle

\begin{abstract}
    The Android operating system runs on the majority of smartphones nowadays.
    Its success is driven by its availability to a variety of smartphone hardware vendors on the one hand,
    and the customization possibilities given to its users on the other hand.
    While other big smartphone operating systems restrict user configuration to a given set of functionality,
    Android users can leverage the whole potential of their devices.
    This high degree of customization enabled by a process called rooting,
    where the users escalate their privileges to those of the operating system,
    introduces security, data integrity and privacy concerns.
    Several rooting detection mechanisms for Android devices already exist, aimed at different levels of detection.
    This paper introduces further strategies derived from the Linux ecosystem and outlines their usage on the Android platform.
    In addition, we present a novel remote rooting detection approach aimed at trust and integrity checks between devices in wireless networks.
\end{abstract}


\section{Introduction}

Privacy and data integrity play an important role in digitized services people use in their every day lives.
When using such services, both the end user and the institution offering the service have great interest in using secure communication
channels and being able to verify the integrity of the applications used.
For those actions, smartphones have evolved to become popular devices to interact with digitized services.
Applications or so-called Apps aim at providing an immersive user experience and integrate deeply within the operating system.
Naturally, the aforementioned privacy and data integrity concerns apply to these applications, too.

The widely used Android operating system~\cite{aosp2020} allows a large range of device vendors to more easily present a variety of devices
all using a common operating system.
This openness has led to a global popularity of Android-based smartphones and other devices using the Android operating system.
It also allows users to gain higher levels of customization of their devices as compared to other competitors.

Institutions offering digitized services provide applications to be used on Android devices.
For some applications, relying on the user to provide device security is sufficient, as no private data may be processed
and no sensitive information is communicated over the internet.
Other applications, e.g. mobile banking applications, are in need of trusted execution platforms.
If data processed by those applications can be obtained by malicious actors,
serious consequences such as identity theft and fraud can occur.
Those applications, but also others having a common interest in data privacy and protection,
have to somehow ensure the integrity of the device they are executed on and the confidentiality of communication channels used.

This can directly contradict with users enjoying the customizability of their devices, as one popular method to
enable broad control over ones' device is the act of gaining elevated privileges through a process called rooting.

As our main contribution we propose a method for remote rooting detection in wireless tethering networks provided by an Android device.
Such networks define a special case of wireless networks, as all communication is routed over the device offering the hotspot,
and can thus be manipulated by said device.
One example for extensive tethering hotspot usage and a matching use case for this remote rooting detection is
the application framework opptain~\cite{ippisch2017infrastructure}.
Additionally, to lay out the options available for applications to secure that the environment they run in allows for processing confidential data,
we provide an overview of rooting methods, corresponding mitigating actions, and present available options for rooting detection.

The remainder of this paper is structured as follows.
We outline other publications related to our work, and reason on the placement of our work within existing research.
Giving definitions of different types of rooting and their impacts, the third section defines the rooting methods we focus our work on.
In the next section, available rooting mitigations and their prospective effects on device usability are stated.
Those measures can impede the rooting process of a device, but as there is no guaranteed security as new exploits can emerge any time,
we list rooting detection strategies commonly used by frameworks deployed on Android devices.
We name additional rooting detection strategies aimed at a variety of rooting techniques using existing functionality included in either the Linux kernel or available applications.
As our main contribution, we then provide a novel strategy aimed at remote root detection on devices connected to a tethering hotspot.
We conclude our paper by summarizing our work and depicting the impact of our contribution regarding existing techniques and rooting scenarios.

\section{Related Work}%
\label{sec:related_work}

The process of privilege escalation and gaining root access on Linux- and UNIX-based devices exists since long before the introduction of the Android operating system.
Methods have been adapted to fit the altered environment provided by Android.
In this section, we give an insight on existing work focussing on rooting and impeding rooting on Android devices.

To give a general overview on available and applicable rooting methods,
several publications such as Yan et al.~\cite{yan2017methods} and Sun et al.~\cite{sun2015rooting} describe a variety of different rooting options for Android devices.

Vidas et al.~\cite{vidas2011all} focus on attacks using rooting methods and their impact on device integrity.
Following their research, having activated rooting techniques on a device leads to an increased risk of further attacks and can enable malicious behavior by other applications.
While we are well aware of this situation, we solely focus on the rooting process on devices without arguing on the usefulness and security impact in general.

With a variety of rooting techniques to choose from, applications aimed at mitigating rooting exploits also have to make use of a large repertory of counter-measures.
This arms race between new rooting techniques and mitigations is detailed in the work of Nguyen et al.~\cite{nguyen2017android}.

Our work provides an overview of the most common rooting detection strategies and the rooting method they focus on.
In addition, we provide further detection strategies enabled by functionality of the Linux kernel,
as well as a novel detection strategy aimed at remote devices connected to a tethering hotspot.


\section{Rooting Techniques}

Rooting an Android device can be done using a multitude of techniques. Some rely
on unwanted behavior of applications or the operating system, while others
follow established ways of gaining higher privileges through tools provided by
the device manufacturer or operating system developer.

For our work, we highlight the two main types of rooting in the following and
detail the differences in how privilege escalation is executed and can be
detected.

\subsection{Definition}
In this section we state a definition of when an Android device shall be called \emph{rooted}.
As Android as an operating system builds upon the Linux kernel, it supports a multi-user concept with multiple distinct user roles and privilege separation.
A super user account, usually referred to as \emph{root}, has advanced permissions granted by the kernel.
Usually, it can override permissions set by any other user and has access to every system function.
Regular apps, on the other hand, should be run from an unprivileged user account, restricting direct access to various functions of the kernel and other functions and files offered by the operating system.

A device is called \emph{rooted}, if an application originally running as an unprivileged user elevates its permissions to those of the super user, e.g., \emph{gained root access}.

\subsubsection{Soft Rooting}
\emph{Soft rooting} is usually not persistent across device reboots.
It could also be fixed through software updates.
Root access gained by soft rooting is hard to detect as there are no changes to the file system.
By investigating the behavior and permissions of currently running processes, those using some sort of soft rooting can be detected.
As this detection mechanism relies on inspecting foreign processes, the detection mechanism itself has to have extended privileges.
Those rooting methods usually rely on a security flaw that can be exploited to elevate permissions.

\subsubsection{Hard Rooting}
\emph{Hard rooting} maintains root access through persistent changes to the file system or a specific partition.
This is usually done by flashing a custom firmware or ROM to the device, voiding the device warranty in most cases.
The most prominent example is Magisk~\cite{magisk}, which only modifies the boot partition, leaving the Android system --- usually located on another partition --- as is.

\subsection{Impact}
Once a device has been rooted, the Android environment has to be seen as compromised.
Previously secure data channels may leak data now~\cite{casati2018dangers}.
Application data can be manipulated, system functions can be replaced, and assumptions on return values and function behavior can differ from what was previously assumed.
This situation creates new challenges for determining the rooting status of a device.

The end user might not be able to identify the device as rooted itself, as previous functionality can be continued unaltered and additional, malicious processes can be run in the background without the need for any user interaction.

\section{Rooting Mitigation}


Most Android smartphone vendors adapt the open-source Android operating system~\cite{aosp2020} to their needs.
This can be necessary to enable device-specific functionality or include vendor branding.
Sometimes, optimizations that should enable sophisticated security functions are introduced, but often the inverse is true as shown in~\cite{projectzero2020securityflaws}.

As proposed by Google's Project Zero, mitigation measures and security functionality included in the Linux kernel should be used instead of “homemade” extensions.
Prominent examples are SELinux~\cite{android-selinux} and AppArmor~\cite{cowan2000subdomain}.
Both strategies are well-established in the Linux ecosystem and can thus be applied on Android-based operating systems with little effort.
Google continues to enable secure default configurations and includes further and recent hardening techniques and tools in newer Android releases~\cite{android_hardening}.

A wide range of vendor specific mitigation measures, both in software and hardware, exist.
While hardware-based measures are hard or impossible to circumvent, software solutions may not be of as high quality as features present in the Linux kernel.
This is due to vendor capacity laid out on software architecture, testing and bug fixing to maintain a valid security measure.

One example of vendor-specific security hardening both in hard- and software is Samsung Knox~\cite{knox}, a platform offering
various security techniques and tools.
As mentioned in~\cref{sec:related_work}, there is an arms race between the development of new rooting techniques and mitigations.
For older versions of Samsung Knox, some attack vectors such as the one mentioned in~\cite{shen2017defeating} exist and
provide well documented instructions for achieving elevated privileges.

Concluding, mitigation measures do not prevent security flaws per se, but can limit the exploitable surface offered through security flaws.

\section{Rooting Detection}%
\label{sec:detection}

With the development of rooting techniques, detection mechanisms have evolved, too.
Applications handling sensitive data, such as financial applications or those associated with the health sector, are in a dire need of verifying the environment they are run in as to not leak any sensitive data to unknown others.

While rooting techniques either work or fail, detecting root access cannot give a distinct answer.
If evidence for achieved root access can be found, the detection mechanism can clearly state that the examined environment has been rooted.
If, on the other hand, no evidence can be detected, the only assumption that can be taken is that no evidence for a rooted environment was found --- it could still be rooted, but not leaving any traces behind that the detection mechanism in use can identify.

We state that this statement is true for all rooting detection mechanisms, as the root account and associated privileges are a part of how the kernel runs the system and will always be existent.
Therefore, a rooting detection mechanism can only search for indications of these privileges made available to a regular user.
These checks run within the possibly rooted environment and can be deluded by other mechanisms of the rooting software.
Examples of such cloaking applications are Magisk~\cite{magisk} and~\cite{rootcloak}.
These applications conceal files, directories and processes most commonly evaluated by root detection applications.
As cloaking applications also only define a predetermined list of actions, rooting detection mechanism applications can advance their detection mechanisms by including the latest indications not yet covered by the cloaking applications.
This leads to an arms race between cloaking and detecting applications as described in~\cite{nguyen2017android}.

In the following, we outline common rooting detection methods already in use by Android applications.
Afterwards, we detail additional sophisticated rooting detection methods and introduce a remote rooting detection strategy.

\subsection{Common Rooting Detection Methods}
The most common rooting detection methods listed below can be found in open source applications as well as in proprietary applications.
While it is straightforward to obtain knowledge of those methods used by open source applications, the methods used in proprietary applications might differ or extend those listed here, as analyzing their processes is more complex and may not even be permitted.

Most rooting detection mechanisms only work for devices using a soft rooting technique, as they rely on changes in the Android environment as indications for their detection mechanisms.
Hard rooting techniques introduce elevated privileges in early stages of the system booting and therefore enable access to otherwise unavailable functionality without leaving traces in the Android environment.

One prevalent open source detection app and library is RootBeer~\cite{rootbeer}.
Its rooting detection methods include those listed in~\cite{netspi2020rootintechniques}, which are part of our outline in the following.

Another framework broadly used on Android devices is Google's SafetyNet~\cite{safetynet}.
This framework is deeply integrated into the operating system and collects information on a devices' state locally.
This information is then submitted to a common backend, where device behavior can be observed in an aggregated way.
Google does not disclose methods used for rooting detection, but we assume at least some overlap with the methods presented below.

\subsubsection{Installed Packages}
Most rooting applications are installed as regular Android packages, i.e. regular apps.
The easiest approach to detect if a device is possibly rooted is accordingly to check if known rooting apps are installed.

\subsubsection{\texttt{su} binary and other installed applications}
The \texttt{su} application commonly found in Linux environments allows the execution of commands and applications with the permissions and privileges of another user.
If such an application can be found, the user can utilize this application to elevate its privileges and gain root access.

Other applications not found on regular Android environments can be installed by the rooting application.
If such an application can be found and the user is permitted to access and execute it, this also indicates the presence of a rooting application.

\subsubsection{Directory permissions}
Most rooting applications modify permissions of system directories, giving access to other users than those permitted regularly.
On regular Android environments, applications should not be able to examine the contents of other applications and processes, but instead communicate through interfaces provided by the environment.
If access to other applications and data belonging to other applications or personal directories for which the user explicitly has to grant permissions can be found, another indication of a rooted environment is found.

\subsubsection{Inspecting available processes}
This method is substantially different to the aforementioned methods, as it can only be employed while a rooting technique is active.
Three possible methods for detecting an active rooting mechanism or manipulated processes are outlined in~\cite{jang2011rooting}.
These methods focus on static checks on application integrity as well as behavior analysis.
For analyzing foreign processes and applications, an elevated privilege level is needed by the application conducting the analysis.
In general, examining differences in device behavior after a rooting attack has been carried out is of greater interest, as it is not time-bound to the moment of the attack.
Most applications in need for rooting detection can only examine the environment during the time they are executed, so the other methods are the most commonly chosen.

\section{Additional Detection Strategies}%
\label{sec:strategies}

\begin{figure*}[t!]
  \centering
  \subfloat[rooted Samsung Galaxy S4]{\includegraphics[width=.33\linewidth]{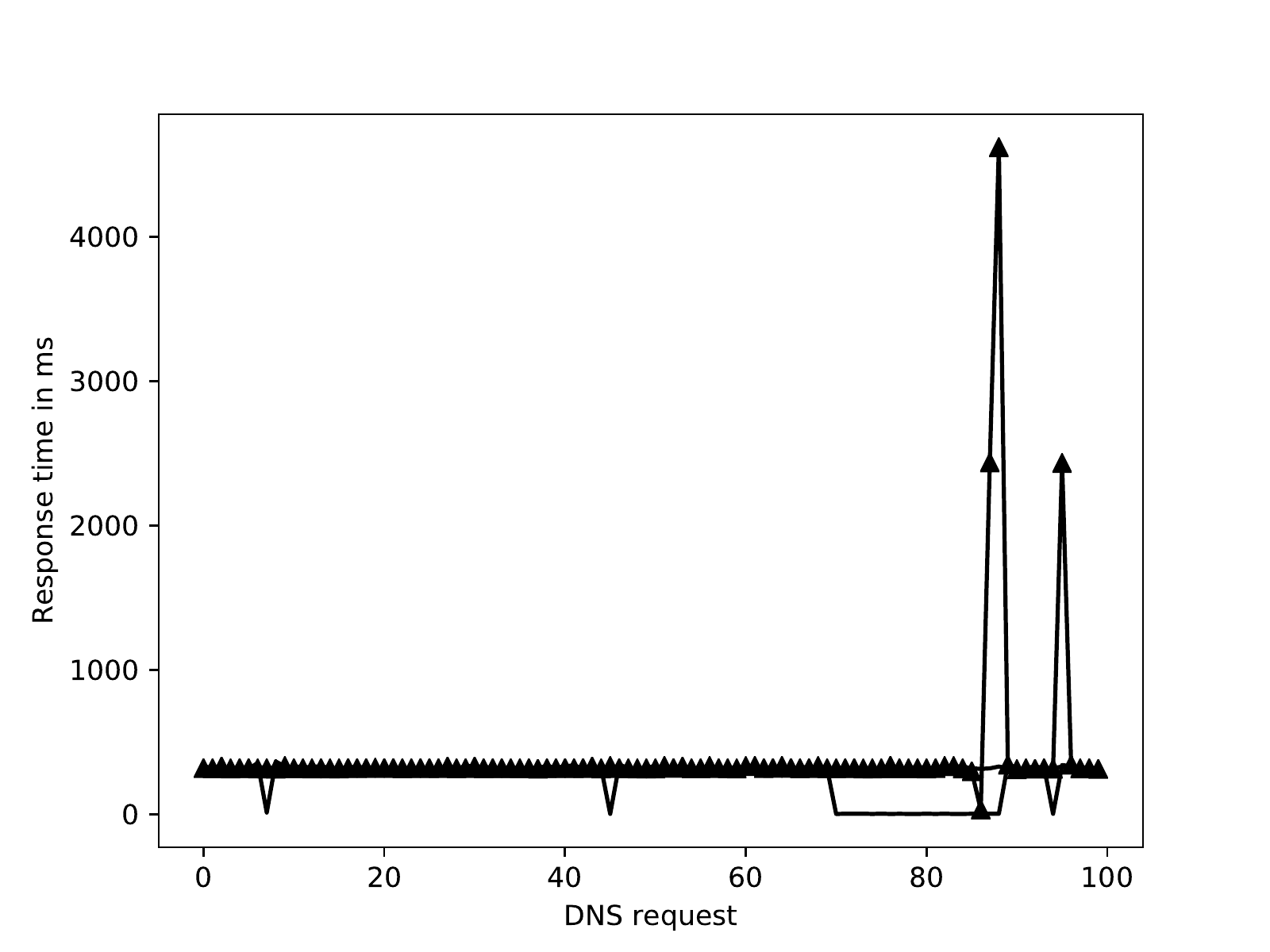}\label{fig:dns_s4_rooted}}
  \subfloat[rooted Samsung Galaxy S5]{\includegraphics[width=.33\linewidth]{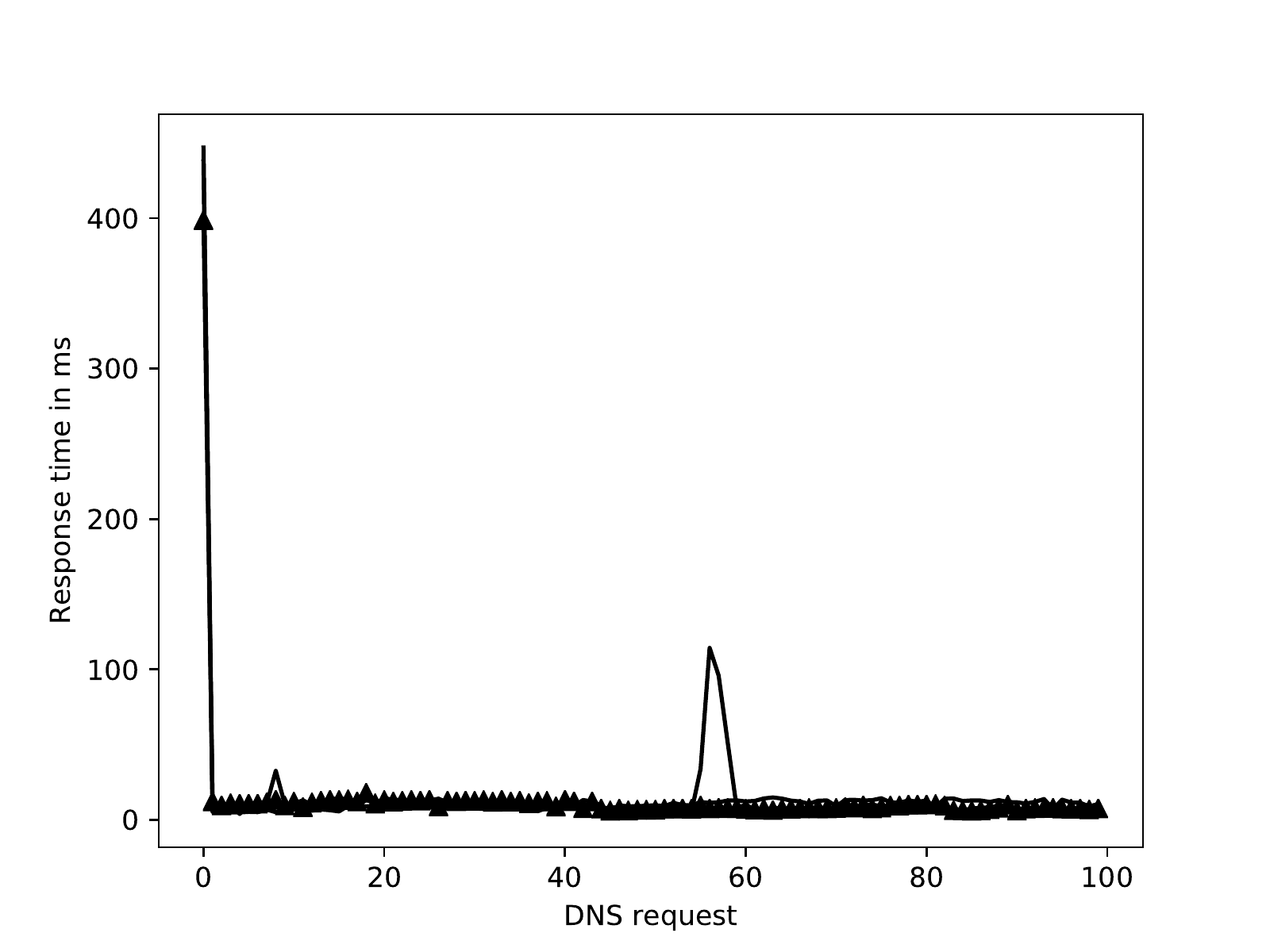}\label{fig:dns_s5_rooted}}
  \subfloat[stock Samsung Galaxy S5]{\includegraphics[width=.33\linewidth]{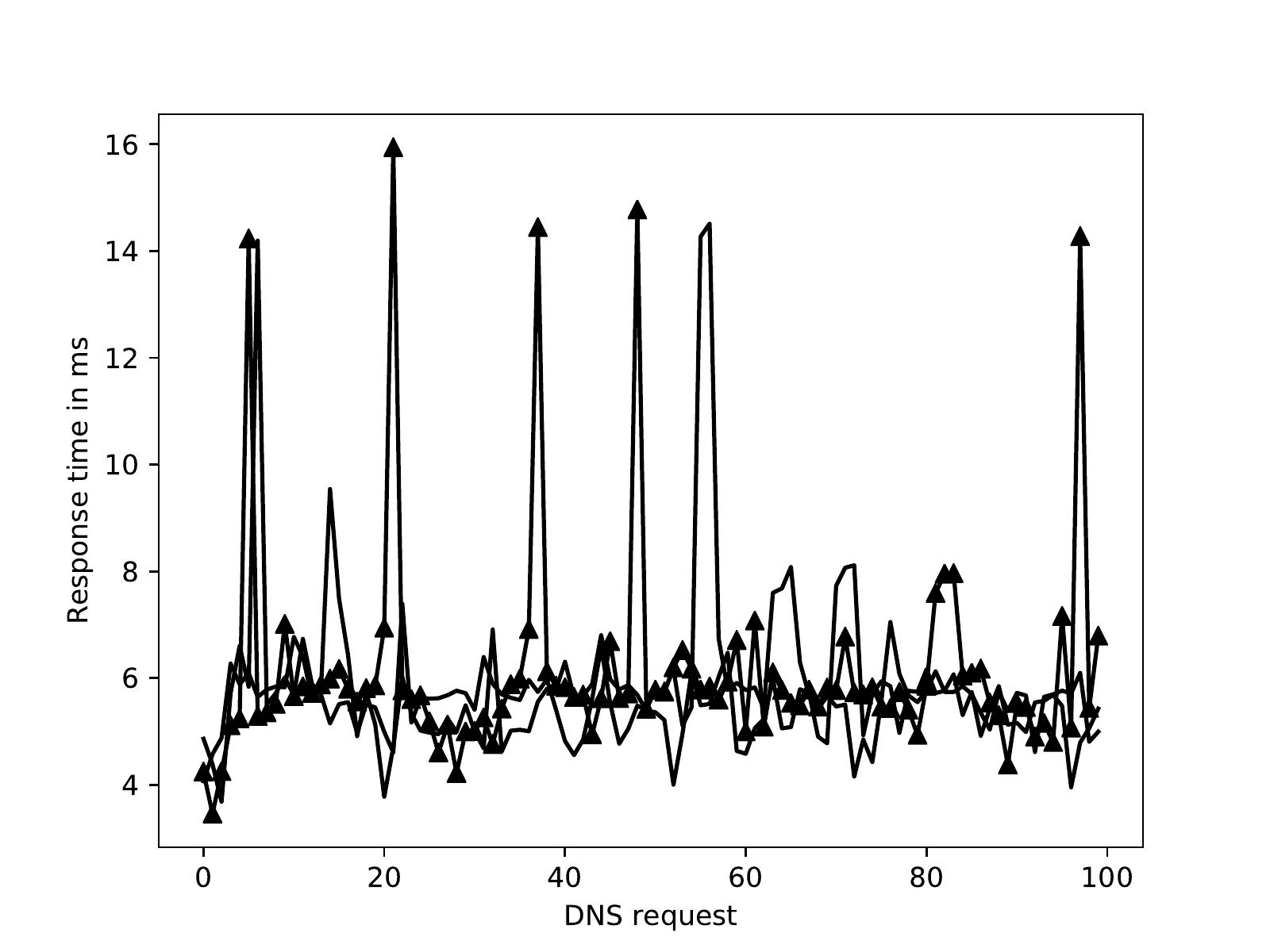}\label{fig:dns_s5_stock}}
  \caption{DNS request and response measurements, three runs of \num{100} samples each}%
  \label{fig:dns}
\end{figure*}

In addition to the detection methods detailed in~\cref{sec:detection}, we conducted research on further detection strategies.
Those strategies build upon existing functionality of the Linux kernel and additional programs and were evaluated for their use within the Android operating system.
In the following we present the method of rooting the given strategy can detect as well as an outline of how the detection strategy works.

\subsection{Detecting custom linked libraries}%
\label{sec:linkedlibraries}

Using library preloading~\cite{ldpreload}, we can inject a custom library into any process we start.
For our tests, we replaced the function \texttt{fopen} as can be seen in~\cref{lst:fopen}.
This function is usually called to open files from the file system.
For our root detection mechanism, we return an invalid call when trying to open the file \texttt{test} and simulate regular function behavior for all other calls.
This test file can then be changed to files known to be related to rooting and further checks and preventions can be introduced at the time of access.
A detailed investigation on the impact of library preloading on Android devices is given in~\cite{evans2015rooting}.

\begin{lstlisting}[float,caption={Function override example for \texttt{fopen}},label=lst:fopen]
FILE *fopen(const char *path,
            const char *mode) {
  if (strstr(path, "test") != NULL) {
    return (FILE *) -1;
  } else {
    FILE *(*orig_fopen)
      (const char*, const char*);
    orig_fopen = dlsym(RTLD_NEXT, "fopen");
    return (*orig_fopen)(path, mode);
  }
}
\end{lstlisting}

\subsubsection{Detection strategy}

We conclude that without further checks, e.g. a signature-based approach for verifying the integrity of shared libraries,
applications handling sensitive data should access critical functions by using statically linked approaches.

\subsection{Detecting process tracing and tampering}

In the previous section, we countered tampering of library
functions which were replaced upon program start. Using
\texttt{ptrace}\cite{ptrace}, a Linux system call to trace processes, a root
cloaking mechanism can intercept every instruction of a traced program and
possibly change instructions or register values during runtime.
This approach enables a broader cloaking methodology, as not only a predefined list of functions and system calls can be intercepted,
but function return values can be inspected and modified to increase the amount of rooting indications successfully cloaked.

\subsubsection{Detection strategy}
To identify if a process is traced by \texttt{ptrace}, the process itself can try to call the \texttt{ptrace} system call on itself.
By definition, only one process can attach and trace a process.
If there is already another process attached, the process itself can evaluate the return value of its own \texttt{ptrace} call to see if it was successful.

Some rooting techniques may also \texttt{ptrace} the init process to tamper with the whole Android environment.
Detecting such attempts can be hard to impossible, as every own attempt to attach to the process to verify its integrity can lead to the process being detected as manipulated by other measures.

\section{Remote Side-Channel Root Detection}

All previously stated rooting detection strategies have to be run on the device assumed rooted.
For some use cases where a device needs to verify the integrity of another Android device it communicates with, those local rooting detection strategies cannot be applied.
From an outside perspective, analyzing any internal information of a remote device relies on information provided by that device.
Verifying the integrity of a remote device therefore relies on additional applications installed on the remote device, which in addition have to assure a secure transmission of integrity check results.
And on top, gaining remote insight on internal information can also be a security risk, as the scope of information gathered has to be restricted so that no access to private or personal information can be obtained.

Our proposed rooting detection strategy focuses on the use case of opportunistic hotspot networking as presented in~\cite{ippisch2017infrastructure}.
To reduce the needed trust and to focus only on the local device, instead we investigated available services of devices participating in a hotspot network, i.e., over Wi-Fi.
Using \texttt{nmap}\cite{nmap}, we scanned the devices listed in~\cref{sec:measurement_setup} for available services.
The results showed that only DHCP and DNS were available for remote devices within the hotspot network.
These two services provide local address distribution and global domain name resolving, both are necessary for participants of the hotspot network to be able to access it for further internet access.
One of the most feasible and non-intrusive methods is the analysis of behavior of those regular services offered by the device.

Multiple measurements can be obtained and evaluated by again different metrics.
As proposed in~\cite{brumley2005remote} and~\cite{brumley2011remote}, time measurements are one promising method of gaining knowledge on a device's internal behavior.
We build our rooting detection strategy upon timing measurements using the available DNS service, which is described in the next section.

\subsection{Measurement Setup}%
\label{sec:measurement_setup}




\begin{table}[tb]
  \centering
  \caption{Average DNS request duration and standard deviation}%
  \label{tab:dns_requests}
  \begin{tabular}{lSS}
    \toprule
    Device & \text{Measured Average (\si{\milli\second})} & \text{Standard Deviation (\si{\milli\second})}\\
    \midrule
    S4 rooted & 258.01 & 131.28\\
    & 404.02 & 520.37\\
    & 318.38 & 32.17\\
    S5 rooted & 16.13 & 43.61\\
    & 13.40 & 38.99\\
    & 14.47 & 45.21\\
    S5 stock & 5.90 & 1.64\\
    & 6.15 & 2.11\\
    & 5.58 & 0.86\\
    \bottomrule
  \end{tabular}
\end{table}

We utilized four devices for our evaluation: Samsung Galaxy S4 and S5, using a rooted and stock configuration each.
Other devices, a Huawei Y3 and HTC Desire 510, were also examined, but no significant difference between rooted and stock versions were observed.
Each device runs a vendor adapted version of Android, which --- besides the rooting on two devices --- has not been modified.
No additional applications were installed as the tethering hotspot can be created using system tools.

For our measurements, we conducted a series of runs for each device separately using the same laptop device as our measurement machine to achieve comparable results.
Also, an environment showing no other wireless networks on regular network scanning was chosen.
Interference between other networks and the device's hotspot network is reduced this way.
All devices have been left running for some time for the following measurements to simulate a \emph{warm start} as compared to a \emph{cold start},
which we will inspect later.

Using the python script%
\footnote{\url{https://github.com/hhucn/android-dns-sidechannel}}%
, we measured the time it took the hotspot device to receive and answer a \texttt{PTR} request for \texttt{8.8.8.8.in-addr.arpa.}.
For each device in stock and rooted configuration, three runs of \num{100} queries each were conducted.

The results are shown in~\cref{fig:dns_s4_rooted,fig:dns_s5_rooted,fig:dns_s5_stock}
and an overview of the averages and standard deviation of each run can be seen in~\cref{tab:dns_requests}.

For the rooted and stock version of Samsung Galaxy S4, no significant difference between our measurements was observed.
Measurements on the rooted Samsung Galaxy S5 show a clear distinction to those on stock configuration.
On the rooted device, DNS query measurements are over two times higher than on the stock device.
We foreclose error introduced by our Python script, as the overhead generated by running the script is consistent between \SIrange{0.3}{0.4}{\milli\second}
and therefore too small to be considered.

Coming back to the aforementioned difference between cold and warm starts of a device, the rooted and stock S5 did not show any measurable difference
when running our script against a freshly restarted device compared to a device that has been left running for some time.
On the S4 on the other hand, the average in measured DNS response delays was more often around \SIrange{4}{6}{\milli\second} on freshly restarted devices,
whereas on a longer running device it occurred in the region around \SI{140}{\milli\second}.
This behavior is consistent on rooted and stock devices and has to be taken into account when deriving information from measurements for root detection.

\section{Conclusion}

In this paper we have shown the motivation and methods behind Android device rooting and noted the security implications introduced by rooting.
An insight on security measures the Android operating system offers, and further frameworks and technologies made available by Google and other vendors was given.
We reasoned that those measures are often not completely enabled due to end user convenience or delays in patch adaption by vendors.
We gave an overview on existing and widely-used rooting detection mechanisms and frameworks, and evaluated additional rooting detection strategies.

As our contribution, we introduced a supplementary remote rooting detection strategy for use in wireless networks Android devices participate in.
This strategy makes no assumptions on available measures on the evaluated device other than default services within a created tethering hotspot being available.
Making it harder for the inspected device to expose the ongoing rooting detection, we showed that through our measurements we can
clearly distinguish between rooted and stock configuration Samsung Galaxy S5.

For general rooting detection, we elect Google's SafetyNet as the most promising framework.
It obtains an integration into the Android operating system which is not available for other apps, as it is integrated by the operating systems' maintainers.
One concern is the submission of locally sourced data to Googles' services for further analysis and detection decision, which occurs in a non-transparent way.

Finally, we conclude that rooting detection might not be a deterministic decision-making process, as assumptions on device environment
may not always hold (e.g., they're manipulated by malicious applications having gained elevated privileges).
The best result one strategy can offer is therefore a tendency between rooted and not rooted, not a binary definite decision.

\subsection{Future Work}
With new mechanisms for rooting Android devices continuously developed as well as counter-mechanisms,
new research input will be available for the foreseeable future.

Remote side-channel root detection, as initially shown by our findings, is one particular topic which can reveal greater use
in the methods of detecting rooted Android devices.
Extending the evaluated scenarios onto further Android devices from a larger variety of vendors has to be conducted in order to
gain deeper insight to general behavior of remotely available services which might be affected by rooting.
We chose DNS queries as our investigated service, while tethering hotspots on Android smartphones offer additional services as DHCP
and hotspot participants can also use different protocols like ICMP Echo Ping.
The relevance of those features has to be examined, too.
We believe that additional conclusions on device integrity can be derived from the analysis of remotely available services.

\section*{Acknowledgments}
The author of this paper would like to thank Dorian Eikenberg for his work on the mentioned and utilized software contained in the Github repository, and conducted measurement scenarios.
\balance{}

\bibliographystyle{IEEEtran}
\bibliography{meine}

\end{document}